\documentclass[12pt]{article}

\usepackage{a4}
\usepackage{epsf}
\usepackage{amsmath}

\begin{document}

\title{Effect of hard processes on momentum correlations}

\author{Guy Pai\'c\\[1ex]
{\it Instituto de Ciencias Nucleares, UNAM, Mexico City, Mexico}\\[1.5ex]
Piotr Krzysztof Skowro\'nski\\[1ex]
{\it CERN, CH--1211 Geneva 23, Switzerland and}\\
{\it Warsaw University of Technology, Faculty of Physics,}\\
{\it ul.~Koszykowa 75, 00-662 Warsaw, Poland}\\[1.5ex]
Boris Tom\'a\v sik\\[1ex]
{\it Niels Bohr Institute, Blegdamsvej 17,}\\
{\it DK--2100 Copenhagen \O , Denmark}}

\date{April 16, 2004}

\maketitle

\begin{abstract}
The effect of hard processes to be encountered in HBT studies at the
Large Hadron Collider have been studied. A simple simulation has allowed
us to generate momentum correlations involving jet particles as well as
particles originating from the kinetic freeze out and to compare them to
a simple theoretical model which has been developed. The first results
on the effect of hard processes on the correlation function for the case
of jet quenching are presented.
\end{abstract}


\section{Introduction}

Momentum correlations of pions and kaons created in heavy
ion collisions have been extensively studied in a wide energy
range from the AGS through SPS and finally at RHIC \cite{hbtreview}. 
The purpose of these studies is  to get information on the 
dimensions of the phase space from which the particles have been 
emitted  \cite{hbtreview2}.  
However, interpretation of the data measured at RHIC 
has been proven to be problematic  and is not yet completely 
understood (see e.g.~\cite{hbtpuzzle}). 
 
Looking ahead towards LHC one can wonder  what are  momentum 
correlations going to look like in presence of an {\em important} contribution 
of hard processes. If we assume that the source is a superposition of 
jet sources and a thermal fireball then the resulting correlations 
should reflect contributions from:
\begin{itemize}
\item pairs of particles from a single jet, which  will reflect  
dimension of the region where the jet fragmented; 
\item pairs of particles from different jets, which will 
be given by the size of the initial collision volume;
\item pairs where both particles stemming from the decoupling of thermal
fireball (larger than the initial collision volume);
\item pairs where one particle comes from the thermal fireball and the other 
one is produced from jet fragmentation.
\end{itemize}

In this paper we describe a toy model to simulate the effects listed above. 
We provide analytical calculations of the correlation function in a special 
limiting case. Then
we investigate how particle production from jet fragmentation shows
itself in the correlation data. We study the visibility  
of the effect at different proportions of jet production to the total 
multiplicity and at different transverse momenta.

\section{Simulation}
\label{simul}

\paragraph{Jets.} 
With Pythia \cite{Pythia} we simulate  p+p collisions
at $\sqrt{s}=5.5\, \mbox{TeV}$.  We extract jets 
in  pseudorapidity coverage of the ALICE detector 
$-1<\eta<1$, by making use of a simple clustering algorithm. 
We identify the highest $p_t$ particle 
among fragmentation products of a single string
as the leading one and then search for particles with an opening angle 
$\Delta\phi$ and relative rapidity $\Delta\eta$ to the leading one that 
fulfills criterion:
$\sqrt{{\Delta\phi}^2+{\Delta\eta}^2}<2$. 
Jets with the sum of $p_t$'s from all particles
between 5 and 10~GeV are selected for further study.

Within the jet, the original position of the particles is 
distributed according to 
a Gaussian of the width $J$ and all particles are assumed to be produced
instantaneously
\begin{equation}
\label{sjs}
s(x;x_j) = \frac{1}{(2\pi\,J^2)^{3/2}}\,
\exp\left ( - \frac{(\vec x - \vec x_j)^2}{2\, J^2}\right )\, 
\delta(t - t_j)\, .
\end{equation}
The jet itself is placed into a random position $(\vec x_j,\, t_j)$. 

\subparagraph{Jet quenching model.}
We model the strong 
jet suppression by a very dense medium by assuming that the jets 
originate from surface of a cylinder with radius $R$ and length $L$.
Eventually, we allow for some finite thickness $\delta R$
for the cylinder surface. It is assumed that the total transverse momentum of 
the jet is always perpendicular to the cylinder surface and that jets 
are produced instantaneously.

\subparagraph{Non-quenching model.}
If the early medium does not suppress jets, we should be able to observe 
jets from hard collisions in all reaction volume. We model such a case by 
choosing a Gaussian distribution of the jets in the transverse plane, but 
we keep the uniform distribution in longitudinal direction from $-L/2$ 
to $L/2$. Instantaneous production is assumed.

\paragraph{Background.}
Apart from jet fragmentation, particles can be produced by  thermal 
fireball. If we are interested in jets, these particles represent  
background for us. Its momentum distribution is such that the {\em total} 
resulting spectrum {\em at low $p_t$} is given by 
\begin{subequations}
\label{bgsc}
\begin{equation}
\label{bgspec}
F(p_t) = \frac{d^2N}{p_t\, dp_t\, dy} \propto \exp\left ( -\frac{m_t}{T}\right )\, ,
\end{equation}

where $T$ is an effective slope parameter. 
The background spectrum is defined as $F_b(p_t)=F(p_t)-F_j(p_t)$,
where  $F_j(p_t)$ the spectrum produced by jets.

{\em At high $p_t$} the spectrum is dominantly shaped by jet fragmentation. 
Technically, it eventually  shows up when $F_j(p_t)$ supersedes parametrisation
\eqref{bgspec}. In order to  obtain a smooth total spectrum with exponential 
low-$p_t$ part and high-$p_t$ tail as given by jets, we choose the 
normalization of the background spectrum $F_b(p_t) = F(p_t)/2$
when the jet contribution becomes $F_j(p_t)>F(p_t)/2$.

The background particles are assumed to be produced instantaneously at 
the same time as the jets and are distributed in space according to a Gaussian
\begin{equation}
\label{bgs}
S_b(x,p) = F_b(p) \frac{1}{(2\pi B^2(p_t))^{3/2}}\, 
\exp\left(-\frac{\vec x^2}{2B^2(p_t)}\right )\, \delta(t - t_0)\, .
\end{equation}
The size of the background source first decreases linearly with $p_t$ up to
$p_t={p_t^{\rm max}}$ and than stays constant:
\begin{equation}
\label{bgb}
B = 
\begin{cases} 
B_0 \frac{p_t^{\rm max}-p_t}{p_t^{\rm max}}
+ B_{\rm max} \frac{p_t}{p_t^{\rm max}}
& \mbox{for}\quad p_t<p_t^{\rm max}\\
B_{\rm max} & \mbox{for}\quad p_t \ge p_t^{\rm max}
\end{cases}\, ,
\end{equation}
\end{subequations}
where we can specify $B_0$, $B_{\rm max}$, and $p_t^{\rm max}$.

\paragraph{Correlation function.}
These are constructed in the so-called {\em out-side-long}
coordinate frame by making use of the weighting algorithm due to 
Lednick\'y \cite{Weights} within the ALICE HBT analysis package
HBTAN \cite{HBTAN}.

\section{Theoretical understanding}
\label{theory}

If we specify the emission function of the source $S(x,p)$, 
correlation function can be calculated from 
\begin{equation}
\label{relcs}
C(q,K) - 1 = 
\frac{\left | \int d^4x\, \exp(i\, q\cdot x)\, S(x,K) \right |^2}{%
\left (\int d^4x\, S(x,K) \right )^2} \, .
\end{equation}
If there are two or more jets in an event, and we sum over 
a large number of events, we can express the emission function by 
integrating over all possible locations of the ``jet centres''
\begin{equation}
S(x,p) \propto \int dx_j\, D(x_j,p)\, s(x;x_j) \, ,
\end{equation}
where $s(x;x_j)$ was specified in eq.~\eqref{sjs} and $D(x_j,p)$ is 
the distribution 
of jet centres, which lead to production of momentum $p$. There may be 
additional dependence of $S(x,p)$ on the momentum, but this is irrelevant
for calculating the correlation function.

For the case of jets produced at cylinder surface we write
\begin{multline}
D(x_j,p) = D(t_j,r_j,\varphi_j,z_j) = \\
\exp\left (\frac{\cos\varphi_j}{\Phi}\right )\,
\delta(r_j - R)\, \Theta(z_j - L/2)\, \Theta(L/2 - z_j)\, \delta(t_j - t_0)\, .
\end{multline}
For simplicity, we neglect the possibility of finite thickness of the surface 
here. The factor $\exp(\cos\varphi_j/\Phi)$ allows particles to be produced 
under some angle with respect to the jet angle. Since we assume that 
the jet momentum is perpendicular to the surface, and we work in the 
out-side-long system, $\varphi_j$ is the angle between $\vec p_t$ of the 
particle and the radial jet coordinate $\vec r$. Parameter $\Phi$ 
regulates
the broadness of the jet: $\Phi \ll 1$ leads to very focussed jets, while 
$\Phi \to \infty$ leaves us with uncollimated particle production.
In principle, $\Phi$ depends on particle momentum, but we will suppress this
dependence here for simplicity.

Correlation function can be calculated from eq.~\eqref{relcs}:
\begin{multline}
\label{ctgen}
C(q_o,q_s,q_l)-1 = \exp\left (- (q_o^2+q_s^2+q_l^2)J^2\right )\, 
\frac{4}{q_l^2\, L^2}\, \sin^2\left (\frac{q_l\, L}{2}\right ) 
\frac{1}{(2\pi)^2\, \mbox{I}_0^2(1/\Phi)}\,\\ \times
\left | \int_{-\pi}^\pi d\varphi\, 
\exp\left [ i\, R\, (q_o\, \cos\varphi + q_s\, \sin\varphi)\right ] \,
\exp\left(\frac{\cos\varphi}{\Phi}\right ) \right |^2\, ,
\end{multline}
where $\mbox{I}_0$ is the modified Bessel function. For qualitative 
understanding it is useful to examine the limit $\Phi\to \infty$ 
when one can analytically calculate the cuts of the correlation function 
along the $q$-axes
\begin{subequations}
\label{ccuts}
\begin{eqnarray}
\label{cout}
C(q_o,q_s=0,q_l=0) - 1 & = & \exp(-J^2\, q_o^2)\,
\left \{ \mbox{J}_0^2(q_o\, R) + \mbox{H}_0^2(|q_o\, R|)\right \}\, , \\
\label{cside}
C(q_o = 0,q_s,q_l=0) -1 & = & \exp(-J^2\, q_s^2)\, \mbox{J}_0^2(q_s\, R)\, ,\\
\label{clong}
C(q_o = 0,q_s=0,q_l) -1 & = & \exp(-J^2\, q_l^2)\, \frac{4}{q_l^2\, L^2}\, 
\sin^2\left(\frac{q_l\, L}{2}\right )\, ,
\end{eqnarray}
\end{subequations}
where $\mbox{J}_0$ is a Bessel function and $\mbox{H}_0$ is a Struve function.

If a background source is added, the correlation function includes 
contributions of pairs of jet particles, background particles and those with 
one particle from a jet and one from background. The analytical expression
thus becomes more complicated. 

Finally, we also made simulations with jets produced from a
Gaussian distribution. Calculations similar to those in this section 
show that such a
distribution of jets leads to a Gaussian correlation function.


\section{Results}

\begin{figure}[t]
\begin{minipage}[t]{6.23cm}
\epsfxsize=6.23cm
\epsfysize=4.5cm
\centerline{\epsfbox{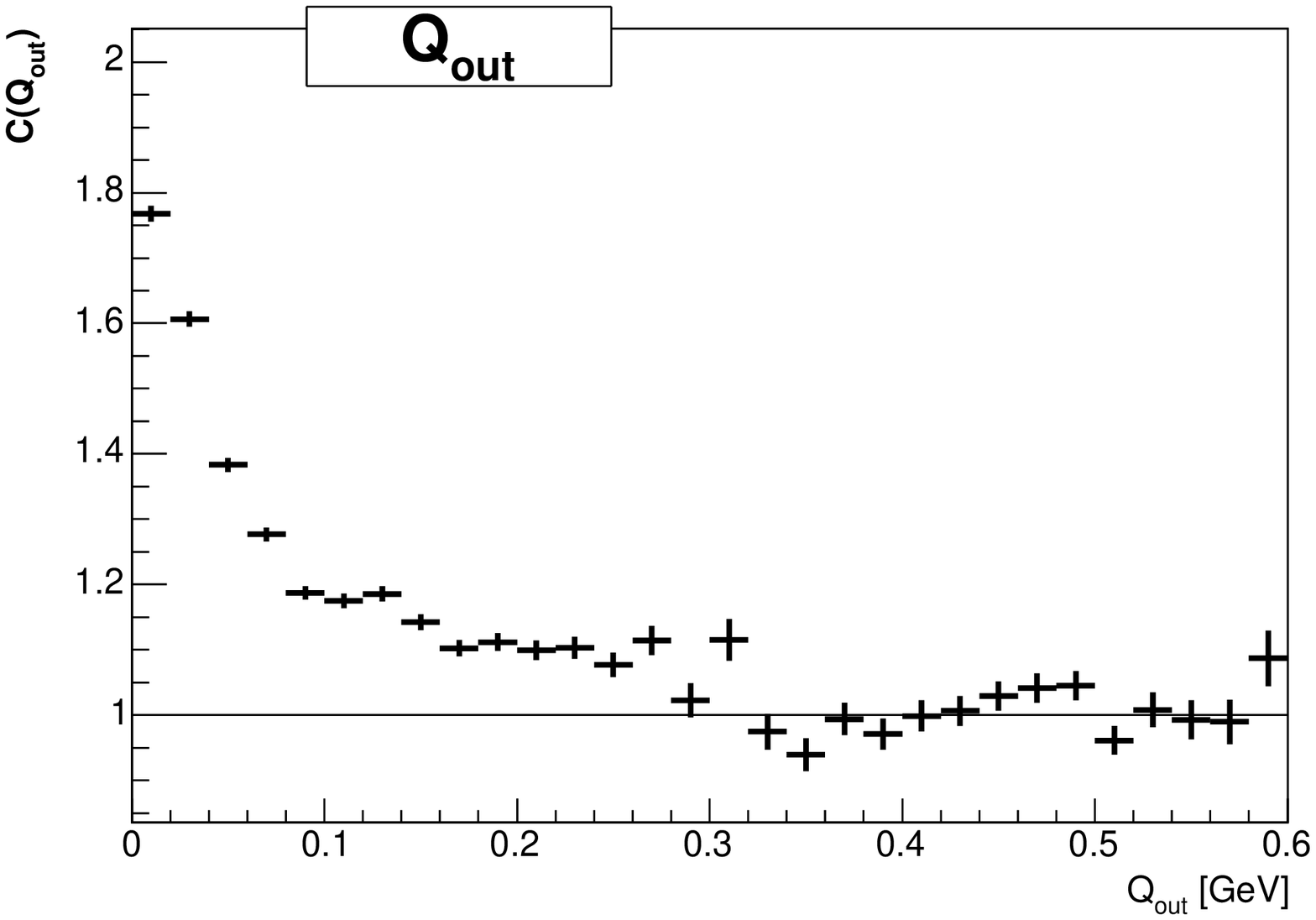}}
\end{minipage}
\begin{minipage}[t]{6.23cm}
\epsfxsize=6.23cm
\epsfysize=4.5cm
\centerline{\epsfbox{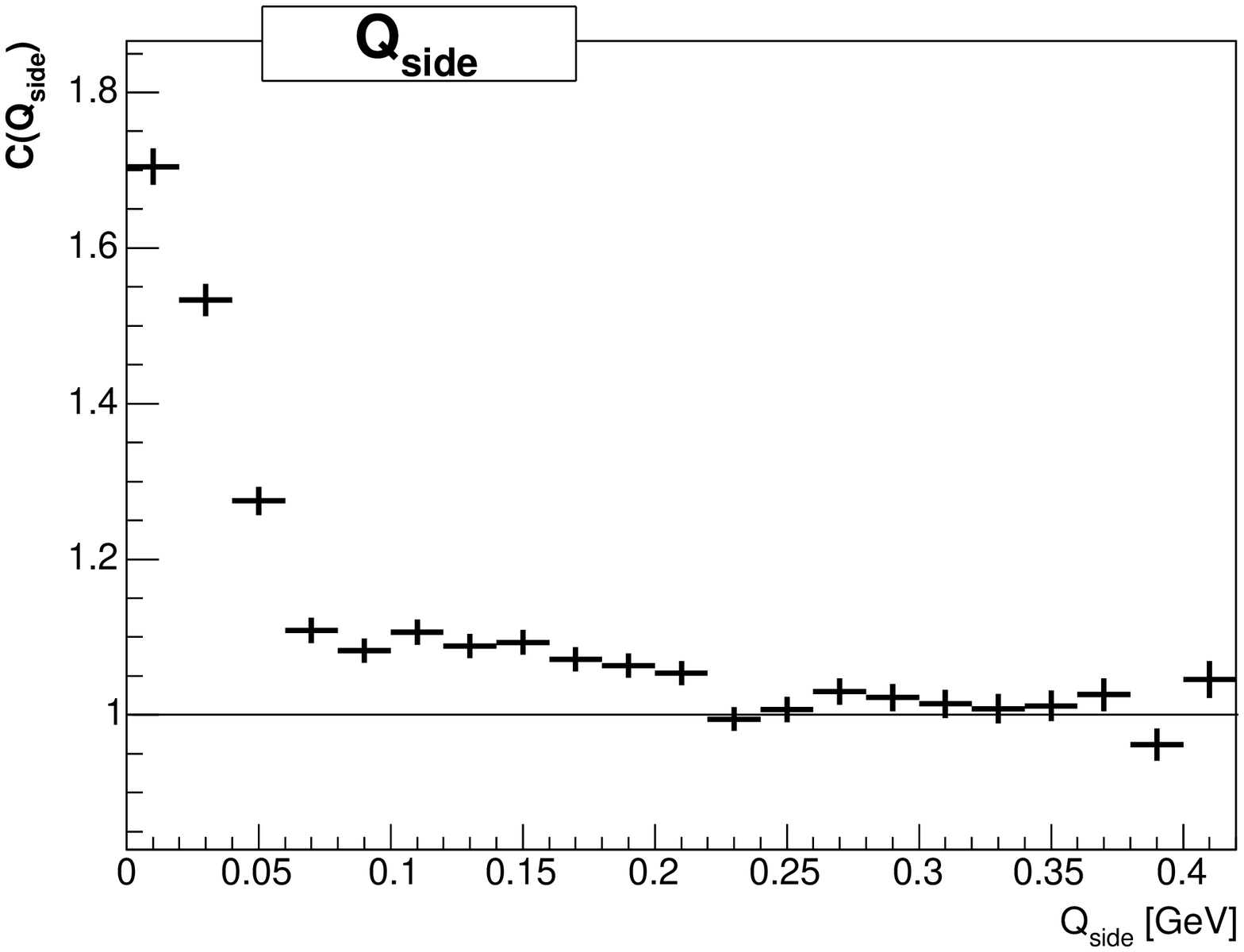}}
\end{minipage}
\begin{minipage}[t]{6.23cm}
\epsfxsize=6.23cm
\epsfysize=4.5cm
\centerline{\epsfbox{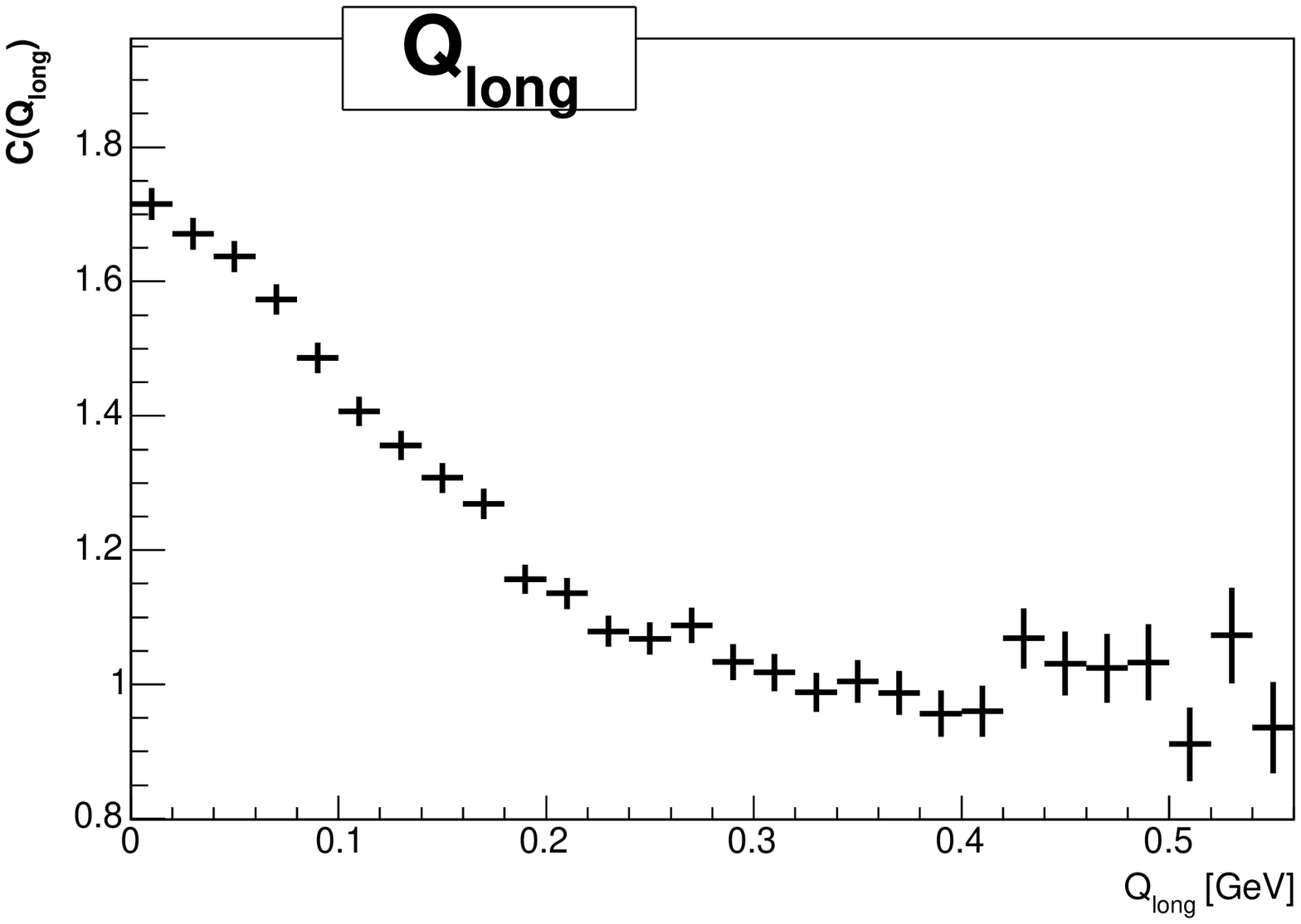}}
\end{minipage}
\caption{
Correlation function resulting from simulation of 10 jets produced at the  
surface of a cylinder with the radius 6~fm and length 1~fm. 
Plotted are cuts along $q$-axes.
\label{f1}}
\end{figure}

\begin{figure}[t]
\begin{minipage}[t]{6.23cm}
\epsfxsize=6.23cm
\epsfysize=4.5cm
\centerline{\epsfbox{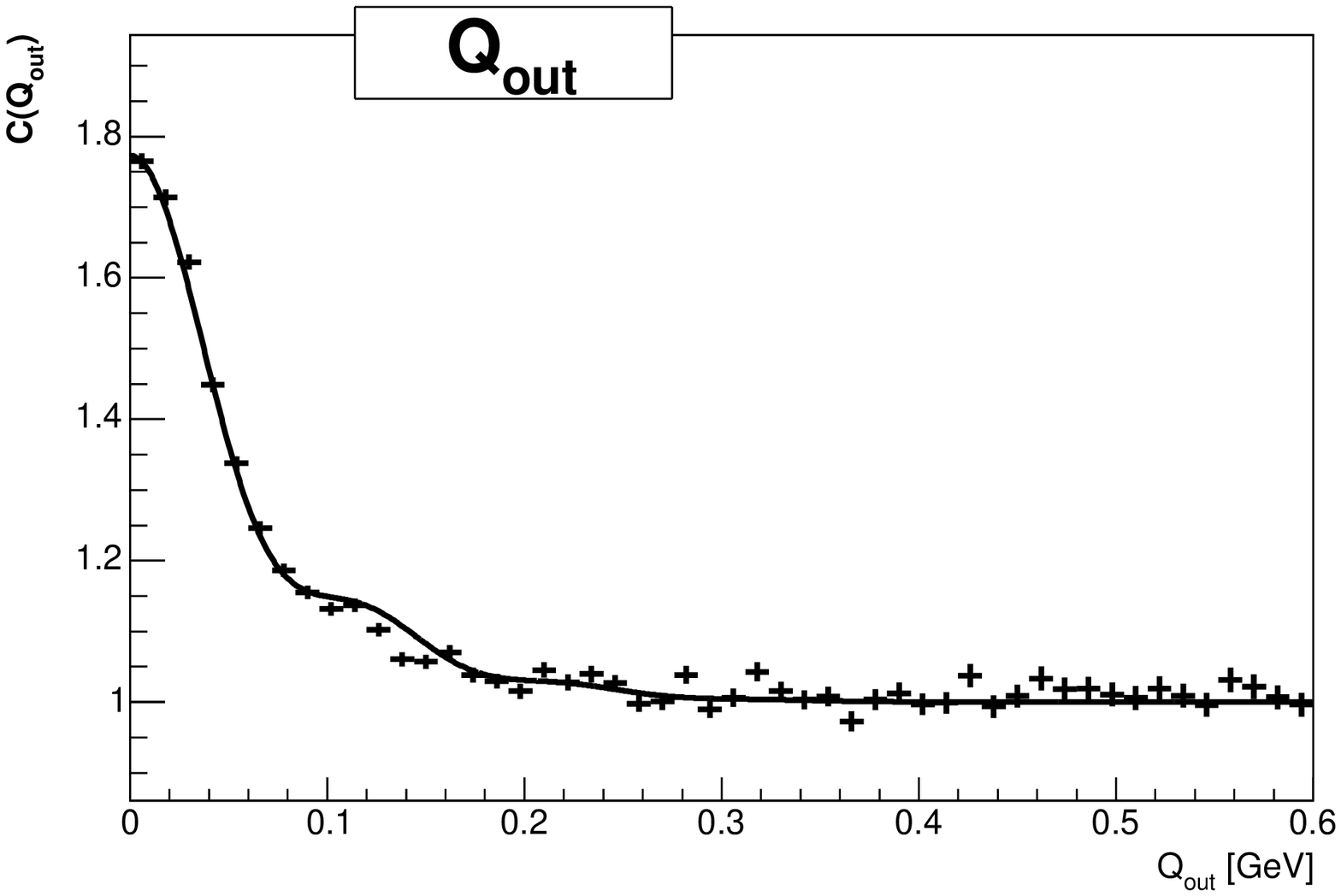}}
\end{minipage}
\begin{minipage}[t]{6.23cm}
\epsfxsize=6.23cm
\epsfysize=4.5cm
\centerline{\epsfbox{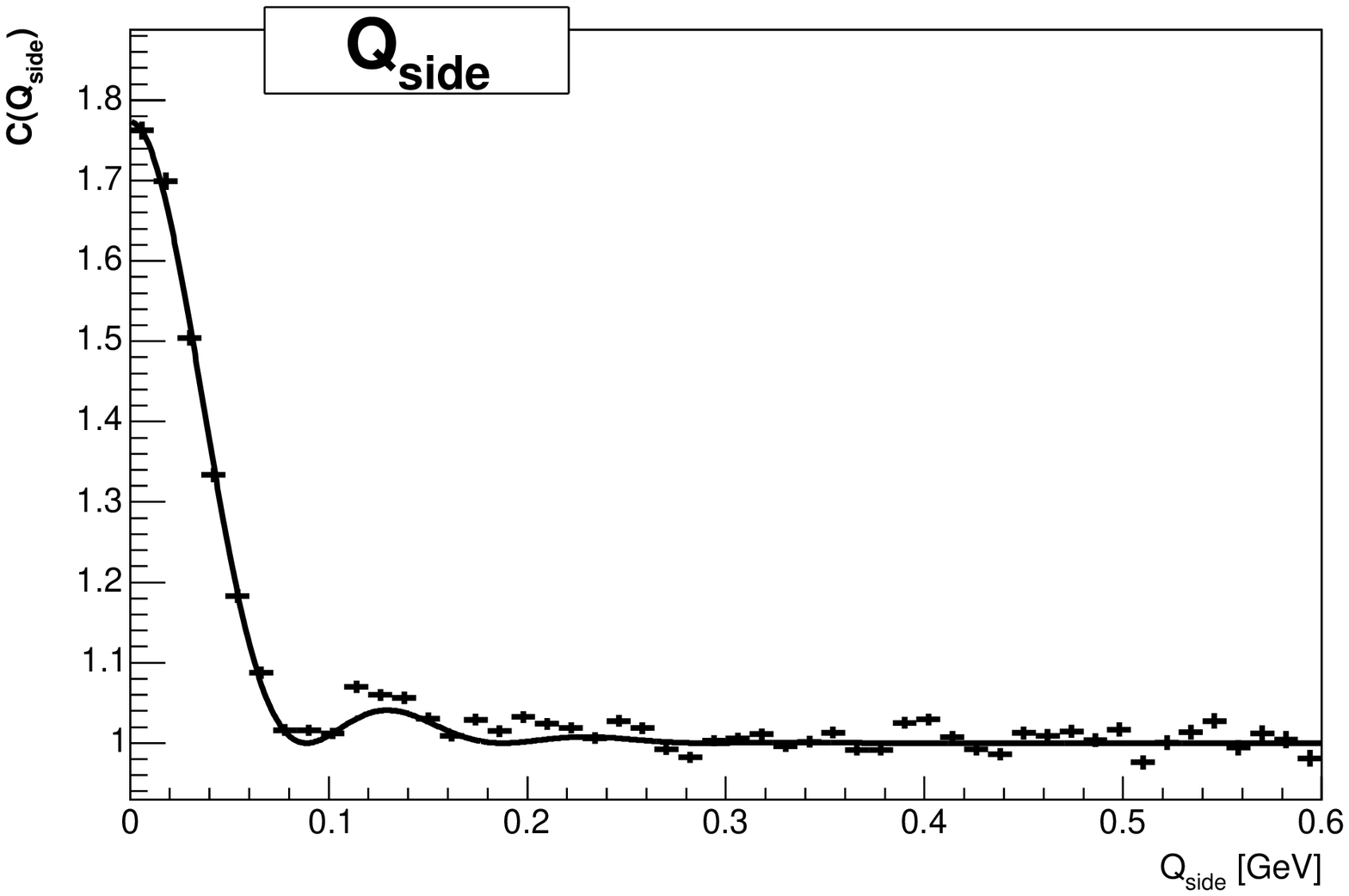}}
\end{minipage}
\begin{minipage}[t]{6.23cm}
\epsfxsize=6.23cm
\epsfysize=4.5cm
\centerline{\epsfbox{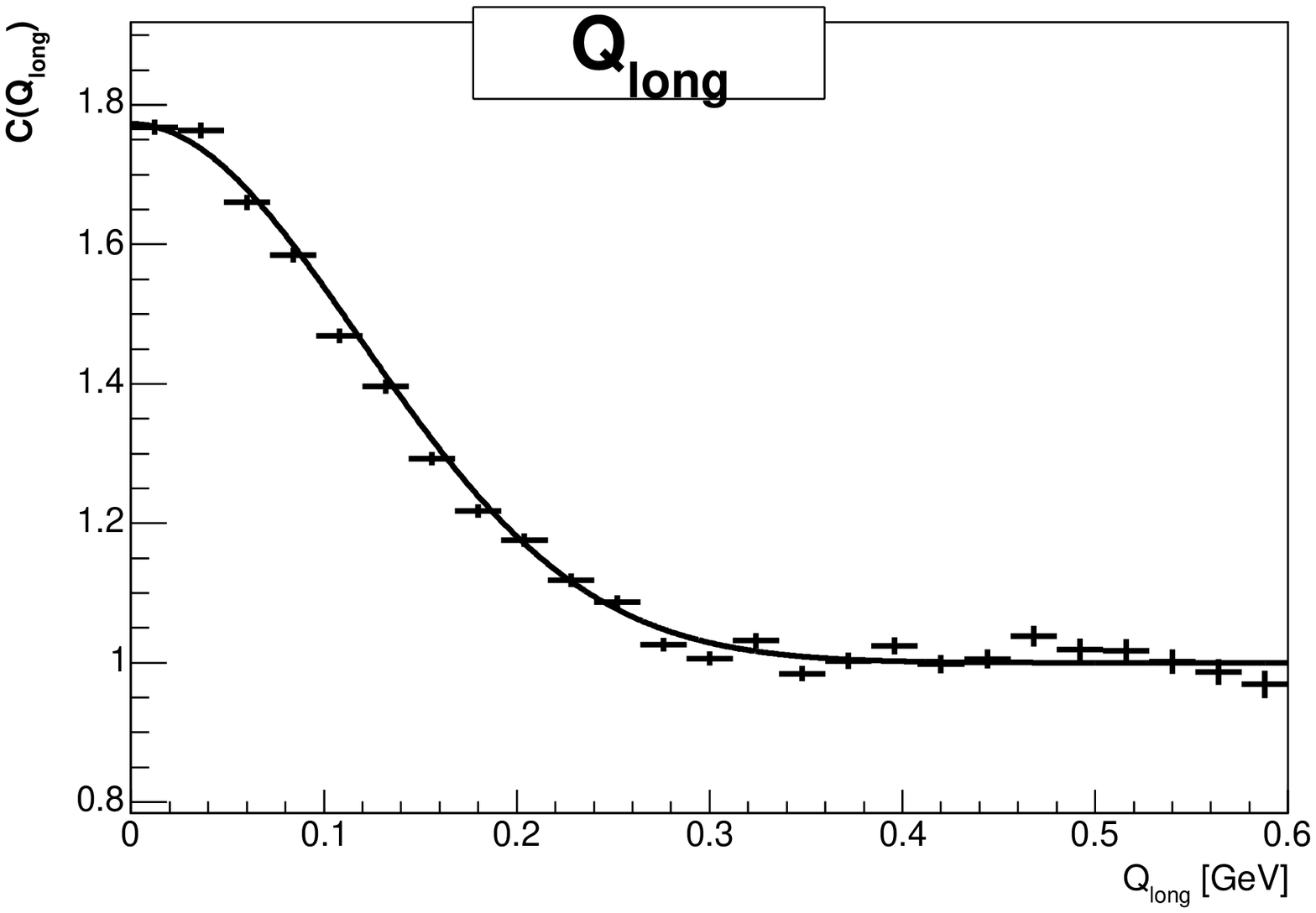}}
\end{minipage}
\caption{
Correlation function resulting from simulation of 100 jets from a surface 
of a cylinder with the radius 6~fm and length 2~fm. Curves correspond 
to a fit with the theoretical curve given by eq.~\eqref{ctgen} multiplied 
by interception parameter $\lambda=0.77$. Parameters of the curves are:
$R = 5.92\, \mbox{fm}$, $J=1.06\,\mbox{fm}$, $L=1.82\, \mbox{fm}$, and
$\Phi = 0.86$.
\label{f2}}
\end{figure}

\paragraph{Only jets.}
First, we analyze events that contain jets only (and no background). We assume 
cylindrical geometry with radius of $R=6\, \mbox{fm}$, 
$\delta R = 1\, \mbox{fm}$, and length $L=1\, \mbox{fm}$ (Fig.~\ref{f1})
or $L=2\, \mbox{fm}$ (Fig.~\ref{f2}). 
Jet radius $J$ is 1~fm.

In Figures~\ref{f1} and \ref{f2} we show 
the projections of the correlation functions from simulations with 10 
and 100 jets, respectively. In the latter case the simulated correlation 
function is rather well reproduced by theoretical curves which were calculated 
from eq.~\eqref{ctgen}. A parameter $\lambda<1$ which multiplies the
theoretical expression was added in order to account for the intercept 
of $C(q=0)$ which is smaller than 2 because in obtaining the projections
we always integrated over finite regions in the remaining two $q$-coordinates. 

The structure of the projection in $q_s$ in Fig.~\ref{f2} can be 
understood from the $\Phi\to \infty$ limit formula, eq.~\eqref{cside}:
the sub-leading peaks stem from the Bessel function. A similar structure
in $q_o$ is ``dissolved'' faster when the jet collimation is finite
and leads to the ``shoulder'' in Fig.~\ref{f2} (left). Projection in 
$q_l$ can be reproduced by eq.~\eqref{clong}.

In the simulation, the angular width of the jet depends on transverse
momentum of particles. Thus $\Phi$ is not constant. It appears that 
averaging over $\Phi(p_t)$ is more important in case of a smaller number 
of jets. Projections in $q_o$ and $q_s$ in Fig.~\ref{f1} result from 
such averaging and we cannot describe them with the prescription
\eqref{ctgen}. Beyond the leading peak we do not see any clear subleading
maxima, but the correlation function assumes
a shoulder-like shape at larger $q$'s.

\paragraph{Jets with background.}
If in addition to jets there is also other  source of particles, 
the characteristic features of the correlation function due to jets
get diluted. We study this by adding a background source, as described 
by eqs.~\eqref{bgsc}. We choose $T=210\, \mbox{MeV}$, $B_0 = 10\, \mbox{fm}$,
$B_{\rm max}=1\, \mbox{fm}$, and $p_t^{\rm max} = 1.2\, \mbox{GeV}$.

\begin{figure}[t]
\begin{minipage}[t]{6.23cm}
\epsfxsize=6.23cm
\epsfysize=4.5cm
\centerline{\epsfbox{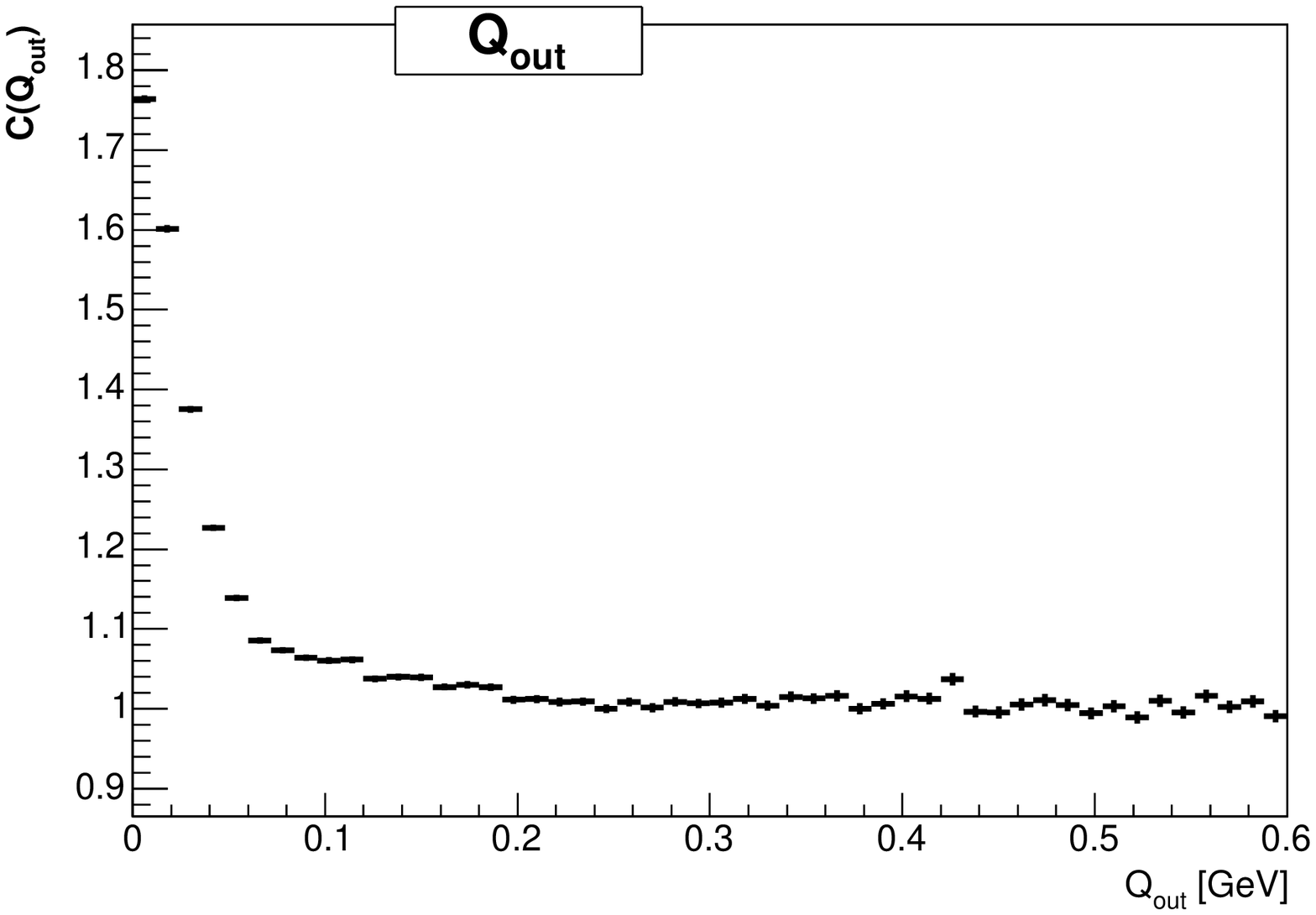}}
\end{minipage}
\begin{minipage}[t]{6.23cm}
\epsfxsize=6.23cm
\epsfysize=4.5cm
\centerline{\epsfbox{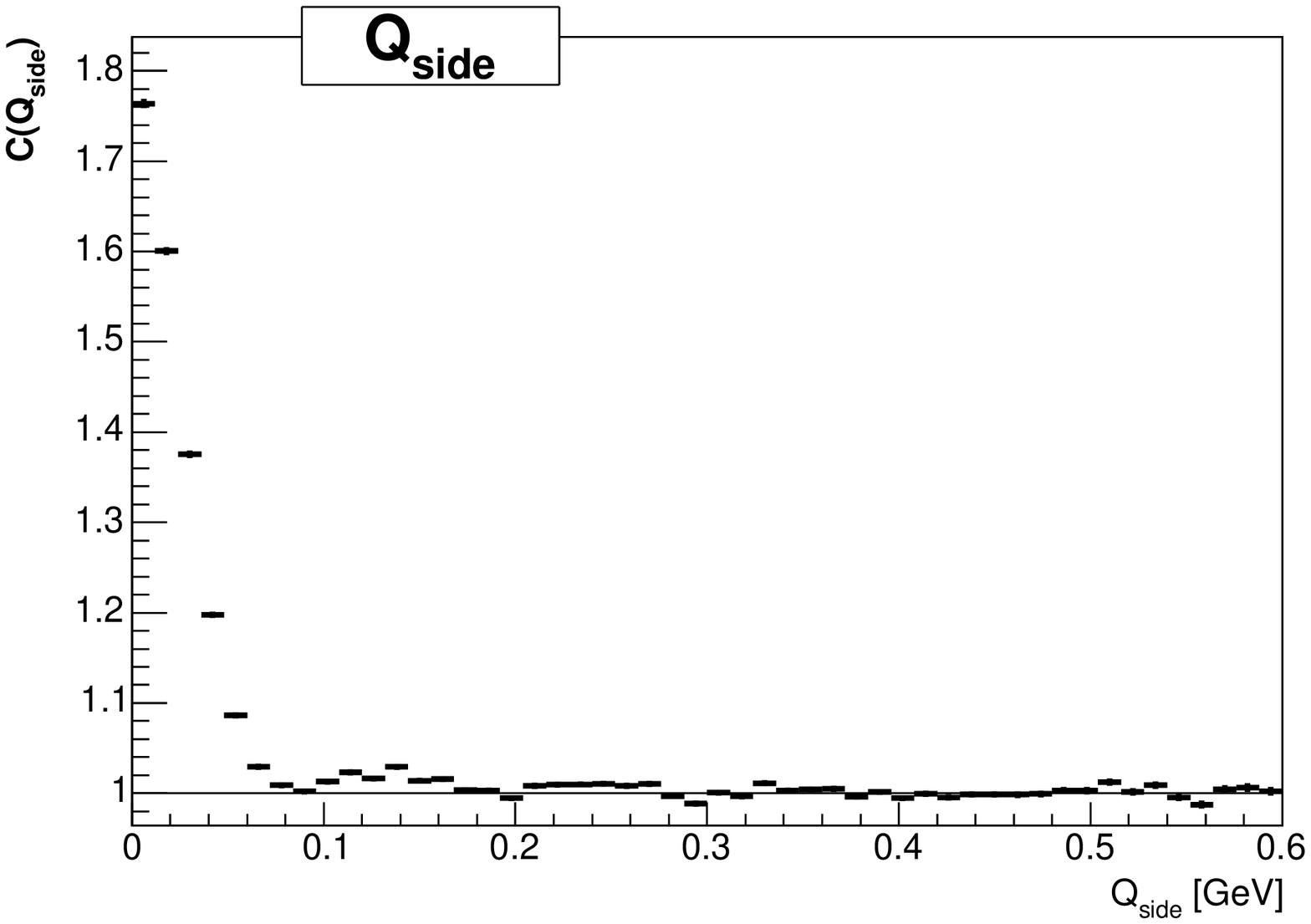}}
\end{minipage}
\begin{minipage}[t]{6.23cm}
\epsfxsize=6.23cm
\epsfysize=4.5cm
\centerline{\epsfbox{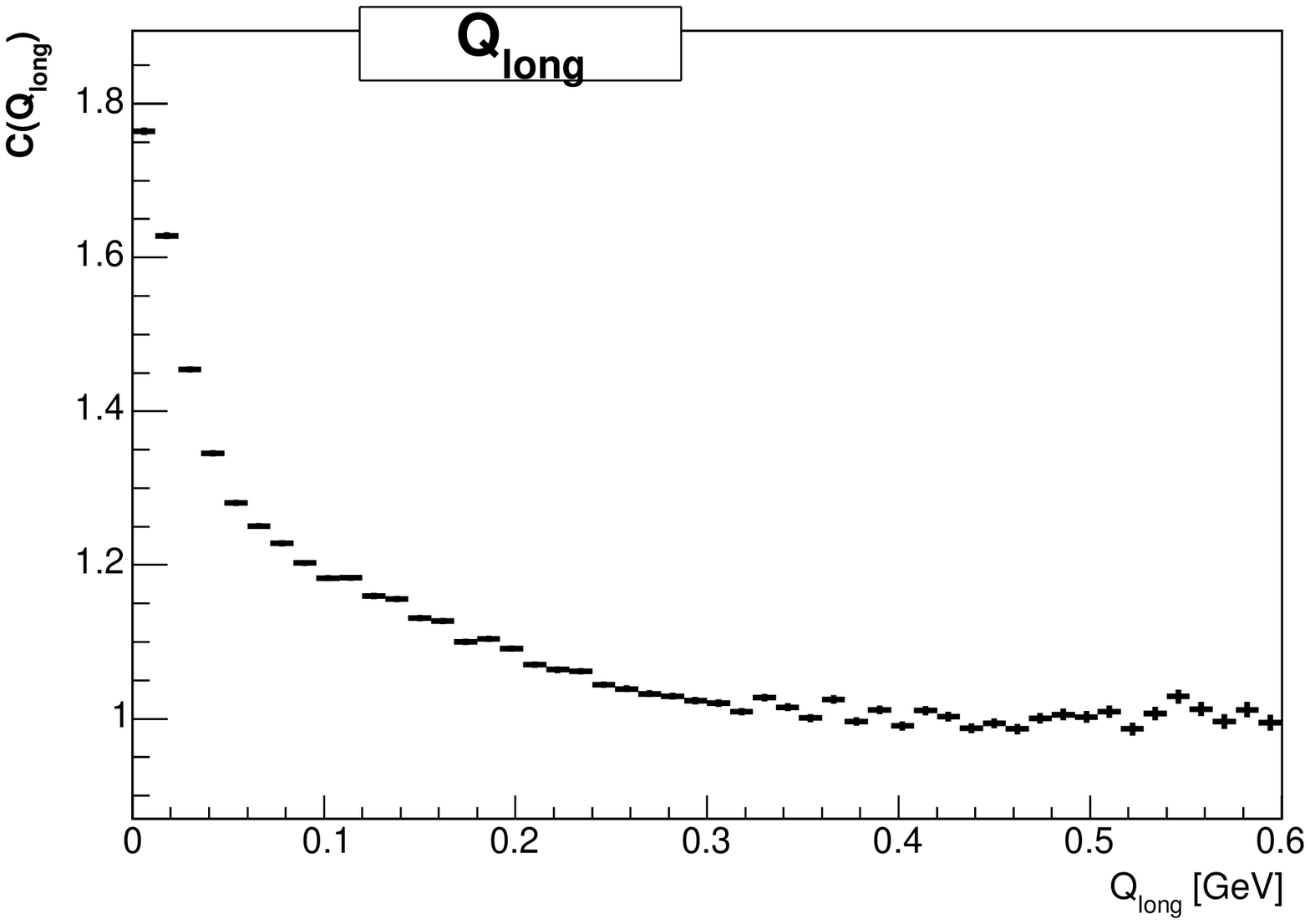}}
\end{minipage}
\caption{
Correlation function resulting from simulation of 100 jets from a surface 
of a cylinder with the radius 6~fm and a ``thermal'' background source; 
2/3 af all pions come from jets. Plotted are cuts along $q$-axes.
\label{f3}}
\end{figure}
In Figure~\ref{f3} we plot the correlation function resulting 
from a simulation with 100 jets where only 1/3 of the pions 
are produced by the background. The dimensions of the background source 
(averaged over $p_t$) are comparable with the transverse size of the source of
jets, while they are clearly larger than it in the longitudinal direction.
Therefore, for the two transverse $q$-components the background just  
weakens the characteristic shape from jets. On the other hand, in 
$q_l$ a narrow peak at $q_l=0$ results from the large size of the background.

We checked that if the percentage of jet pions drops as low as 1/6,
the $p_t$-integrated 
correlation function is completely determined by the background source and 
no signal of jets is seen. Note that the proportion of jet particles  
can be even lower at ALICE: at midrapidity we assume 3000 for $dN/dy$ of 
charged particles, out of which about 90\% are pions, so we obtain
roughly  1350 single-charge pions. Since a jet produces on average 2.4 positive
pions in our acceptance region, 100 observable jets with pseudorapidity
$-1 < \eta < 1$ will leave us 
with 120 single-charge pions per pseudorapidity unit. 
That makes the ratio of pions from jets
to the total pion number about $120/1350 \sim 1/11$.
(We explore this situation in the next paragraph.)

\paragraph{High $p_t$.}
\begin{figure}[t]
\begin{minipage}[t]{6.23cm}
\epsfxsize=6.23cm
\epsfysize=4.5cm
\centerline{\epsfbox{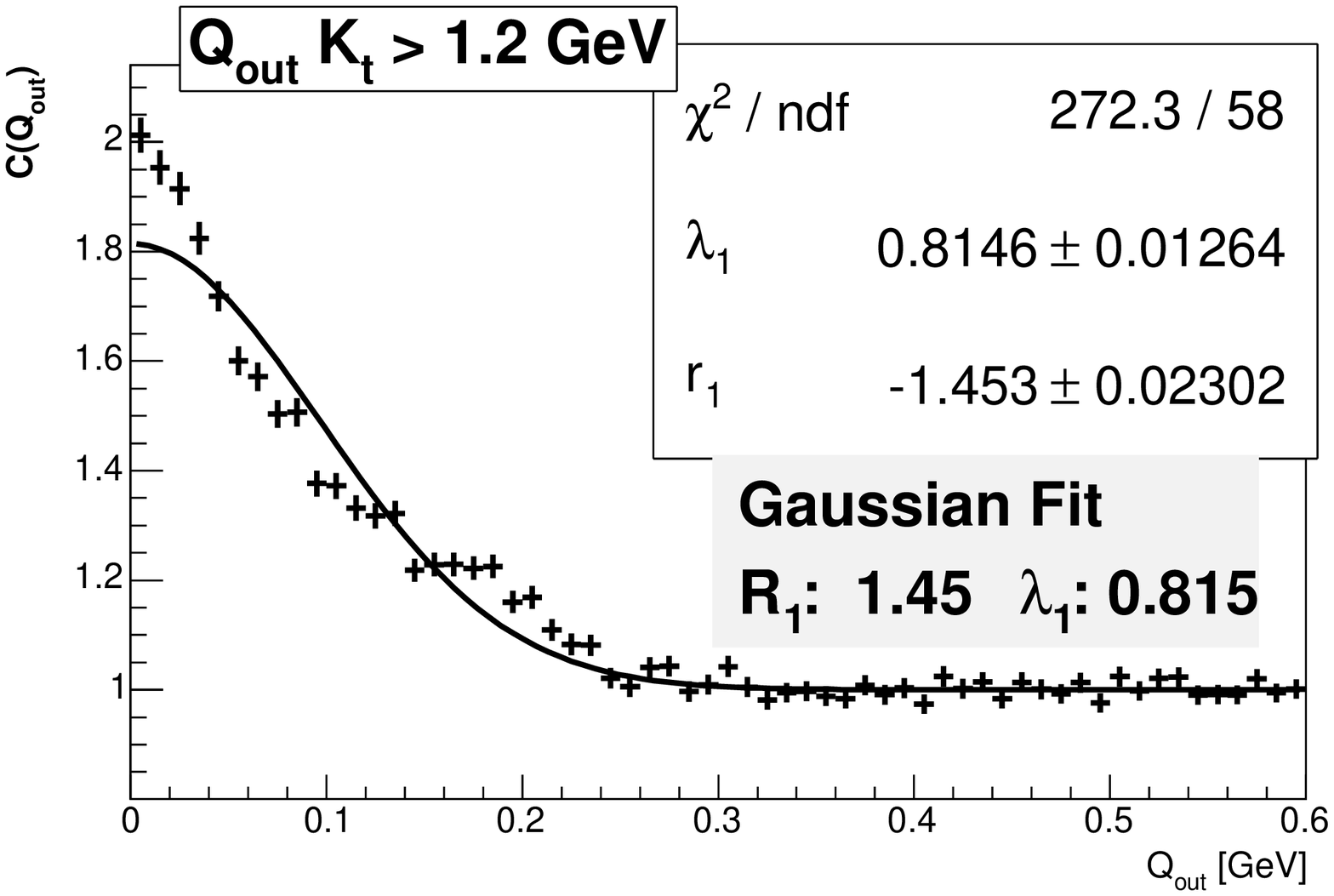}}
\end{minipage}
\begin{minipage}[t]{6.23cm}
\epsfxsize=6.23cm
\epsfysize=4.5cm
\centerline{\epsfbox{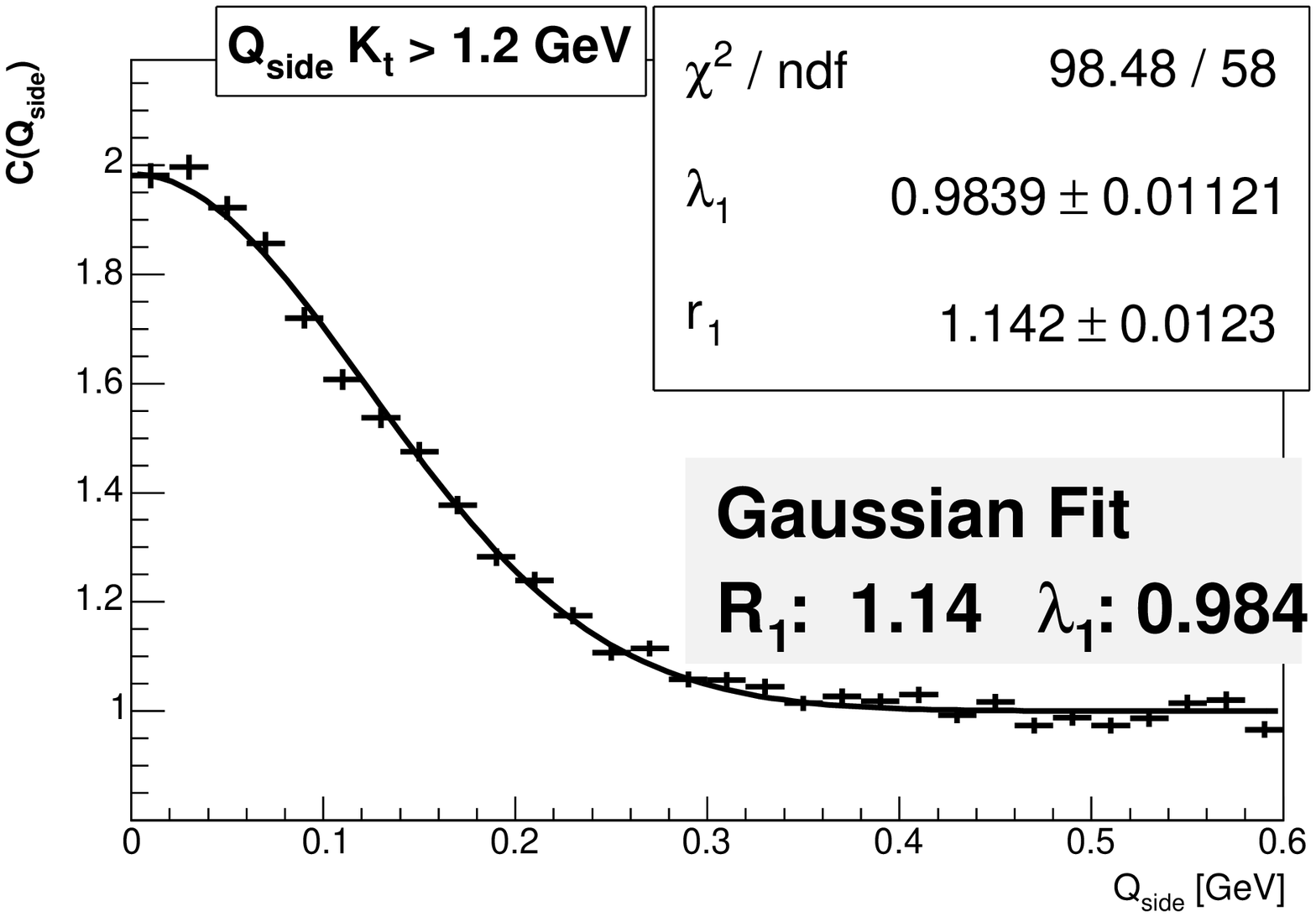}}
\end{minipage}
\begin{minipage}[t]{6.23cm}
\epsfxsize=6.23cm
\epsfysize=4.5cm
\centerline{\epsfbox{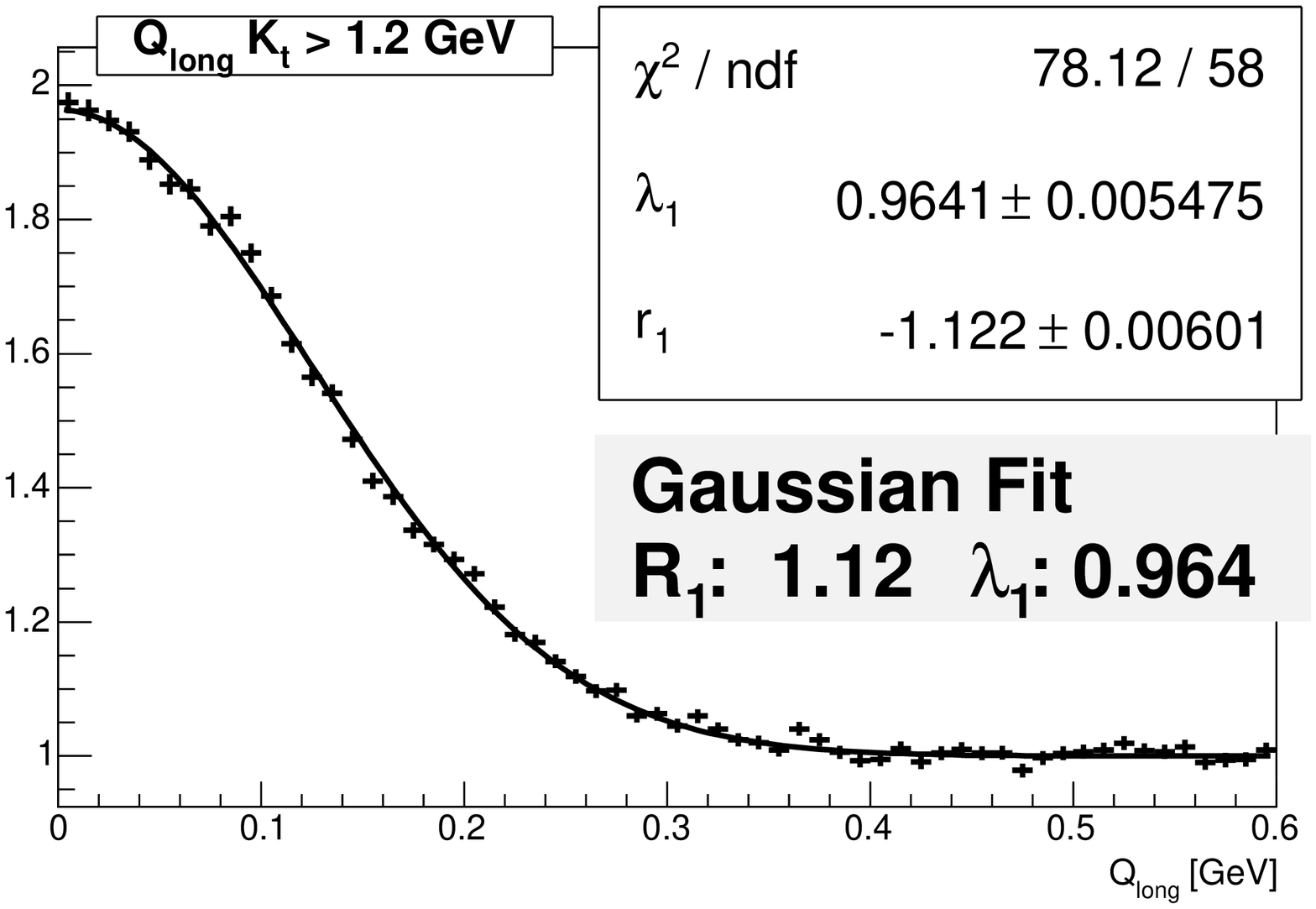}}
\end{minipage}
\caption{
Correlation function resulting from simulation of 100 jets from a surface 
of a cylinder with the radius 6~fm and a ``thermal'' background source; 
1/12 of all particles come from jets and a cut is imposed: 
$K_t>1.2\, \mbox{GeV}/c$. Plotted are cuts along $q$-axes together with 
Gaussian fits.
\label{f4}}
\end{figure}
In order to eliminate the effect of the background we can focus
on particles with high transverse momenta where we expect the 
portion of jet-produced pions to be increased. In Figure~\ref{f4}
we plot correlation function calculated for jets with a background 
source as previously, but we use only particle pairs with $K_t$ above 1.2~GeV
in our analysis.
A non-Gaussian shape, particularly in $q_o$ and slightly in $q_s$,
is observed, which is due to 
contribution from jets. 

We have also simulated jet production from a source with Gaussian
transverse profile together with background 
and imposed the same contraint on $K_t$ (not shown). This  leads 
to similar non-Gaussian shapes of the correlation function as shown
in Fig.~\ref{f4}.
A more detailed study of the shape of the correlation function
is neccessary in order to understand  differences between the 
two models for jet distribution.

\section{Conclusions}

\begin{itemize}
\item Correlation functions are affected  if 
there are pions produced from fragmentation of jets. 
\item Correlation functions assume particularly characteristic shape if
there is strong suppression of jets due to energy loss
of the leading parton in a deconfined medium.
\end{itemize}
In the latter case, jets are effectively emitted only from 
the surface of the fireball.
Presumably, these characteristic shapes are dissolved in real data
because there are much more particles stemming from thermal fireball.

Correlation signal of individual jets is not visible if the total number of
particles in the bin is much larger than the 
number of particles produced by a jet.
We can estimate the contribution to each bin of correlation function 
as $N_{tot}^2$ ($N_{tot}$ is total number of particles).
We write $N_{tot} = N_j+ N_b$, where $N_j$ is the number of 
particles produced by jets and $N_b$ the number of particles 
coming from the thermal source. Then the number of pairs
is written as 
$(N_j+ N_b)^2 = {N_j}^2 +2N_jN_b+ {N_b}^2$. In our study, 
${N_j}^2$ corresponds to signal, ${N_b}^2$ to background, 
and $N_jN_b$ is the correlation between particles of signal and
background. If $N_j \ll N_b$, the signal  becomes invisible.

\begin{itemize}
\item
The influence of background can be limited by using only pairs with 
high transverse pair momentum $K_t$ in the analysis.
\end{itemize}
Although we couldn't describe the correlation 
function obtained after such a cut-off by our analytical expressions,
we could observe a clear deviation from Gaussian shape. Note that 
Gaussian shape  would result from our background source in the absence of
jets. A  non-Gaussian shape can influence 
results of fitting  real data by Gaussian parametrisation and may 
cause problems when interpreting such a fit. We have shown that jets
lead to structures in the correlation functions at large momentum differences:
beyond the main peak at $q=0$. Thus it is also important to look at the 
large-q region in order to learn about the characteristics of the source.


\end{document}